\documentclass[fleqn,usenatbib]{mnras}

\usepackage[T1]{fontenc}
\usepackage{ae,aecompl}

\usepackage{newtxtext,newtxmath}

\usepackage{graphicx}	
\usepackage{amsmath}	
\usepackage{tablefootnote}	

\title[Formation of galactic bulges]{Formation of galactic bulges from the cold gas filaments
in high-redshift dark matter halos}

\author[M. Noguchi]{
Masafumi Noguchi $^{1}$\thanks{E-mail: noguchi@astr.tohoku.ac.jp (MN)}
\\
$^{1}$Astronomical Institute, Tohoku University, 6-3, Aramaki, Aoba-ku, Sendai, Miyagi, 980-8578, Japan
}

\date{Accepted XXX. Received YYY; in original form ZZZ}

\pubyear{2020}

\begin{document}
\label{firstpage}
\pagerange{\pageref{firstpage}--\pageref{lastpage}}
\maketitle

\begin{abstract}

	Formation process(es) of galactic bulges are not yet clarified although
	several mechanisms have been proposed. In a previous study, we suggested
	one possibility that
	galactic bulges have been formed from the cold gas inflowing through 
	surrounding hot halo gas in massive dark matter halos at high redshifts. 
	It was shown that this scenario leads to the bulge-to-total 
	stellar mass ratio increasing with the galaxy mass, in agreement with the 
	well-known observed 
	trend. We here indicate that it also reproduces recent observational results 
	that the mean stellar age of the bulge increases with the galaxy mass while 
        the age gradient across the bulge decreases.
	We  infer that this formation path applies mainly to high-mass galaxies 
	and the bulges in lower-mass galaxies 
	have different origins such as secular formation from the disc material.

\end{abstract}




\section{introduction}

Assembly of galaxies in the early universe is a matter of intense debate in current astrophysics. Among others, the formation of bulges is a key ingredient which brought about the 
morphological diversity of the present-day disc galaxies.
Despite much effort in clarifying the bulge formation process from both observational
and theoretical perspectives, we are still far from satisfactory understanding 
of this important piece of galaxy formation.
Complex structures and kinematics of galactic bulges, especially the dichotomization
into classical and pseudo bulges,
suggest contribution of 
several mechanisms in their formation process.
\citep{ko04}.
Classical bulges are usually linked to early formation by direct collpase \citep{la76,zo15} and/or
minor galaxy mergers \citep{ho09}. 
Pseudo bulges are often allged to be the product of the secular formation processes
from the disc material, such as gas infall induced by galactic bars 
 \citep[e.g.][]{at92},
the bending instability of the bars themselves
 \citep[e.g.][]{ra91}, and inward migration of massive clumps formed in gas-rich 
 young galactic discs
 \citep[e.g.][]{no98,no99,in12,bo08}.

Any consistent theory of bulge formation must explain the observed properties of other 
galactic components in the same framework.
In seeking such a picture, we are working on the galaxy evolution model based on 
the cold accretion picture for gas accretion onto forming galaxies
 \citep[e.g.][]{fa01,ke05,de06}.
 \citet{no20} suggested a new possibility that the bulge formation is fueled 
  by the cold gas streams 
characteristic of the halo gas in massive galaxies at high redshifts.
It was found that this picture can reproduce the observed trend that 
the mass fraction of the bulge relative to total stellar mass of the galaxy 
increases with the galaxy mass.
We here report that the same model can also explain the observed age structures of
galactic bulges, namely the mass dependence of 
the mean stellar age and  the age difference within the bulge region.

\begin{figure*}
	\includegraphics[width=0.6\linewidth]{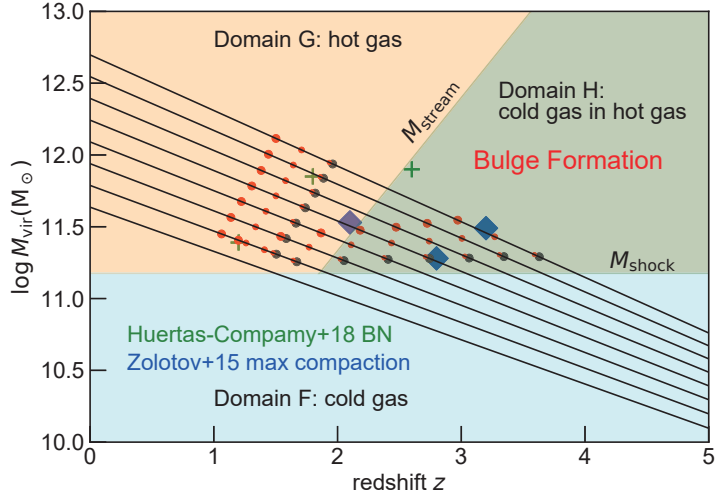}
    \caption{
   Evolution of the virial mass is indicated by sold lines for
      eight models analyzed in the present study overlaid on the three domains
	for the different gas states.
	Black circles on each evolution path indicate the two epochs bewteen which 
	the cold gas in Domain H arrives at the disc plane.
	Red circles indicate, in increasing size, 
	the times $t_{\rm arr}+t_{\rm dyn}, t_{\rm arr}+10t_{\rm dyn}$,
	and $t_{\rm arr}+20t_{\rm dyn}$.
	Here, $t_{\rm dyn} \equiv (G M_{\rm gal}/R_{\rm gal})^{1/2}$, where the galaxy radius is
	set to be $R_{\rm gal} = 0.1 R_{\rm vir}$ considering the high spin parameter $\sim 0.1$
	at high redshift \citep{da15} and the galaxy mass $M_{\rm gal}$ includes all stars and
	the portion of dark matter within the galaxy radius.
	The peak masses for the blue nuggets observed by \citet{hu18} for three 
	different redshifts are shown by green crosses, whereas the blue diamonds  
	are the maximum compaction in three galaxies in VELA simulation by \citet{zo15}.
	The virial masses for the observed BNs are derived by using the stellar-to-halo
	mass ratio (SHMR) by \citet{ro17} for the corresponding redshifts. The virial masses 
	for the simulated galaxies are extrapolated along the expected evolutionary tracks 
	(black lines) from the cited values at $z=2$.
	\label{Fig.1}}
\end{figure*}

\section{Models}

The cold accretion theory has been proposed on the basis of realistic simulations for
thermal and hydrodynamical evolution of the primordial gas in the cold dark matter universe
 \citep[e.g.][]{fa01,bi03,ke05,de06,oc08,va12,ne13}.
 It states that
 the intergalactic gas flows into the hierarchically growing dark matter halos in unheated state 
 and fuels the forming galaxies
 except in the most massive halos in recent cosmological epochs, where the cooling flow
 of the shock heated halo gas prevails.
 This picture represents a major modification to the long-standing paradigm
 which aruges that the heating by shock waves is the universal 
 behaviour of the gas that enters growing dark matter halos
     \citep[e.g.][]{re77}.
This new scenatio provides possible solutions to several 
observations unexplained in the shock-heating theory, including
the existence of abundant luminous galaxies at high redshifts 
and very red colors (and therefore complete quenching of star formation) 
of present massive elliptical galaxies.
     \citep[e.g.][]{ca06,de09}.

Recently, the application of this scenario was extended to subgalactic scales.
\citet{no20} examined the morphological buildup of disc galaxies under the cold accretion while
  \citet{no18} tried to explain the chemical bimodality
  observed in the Milky Way disc stars
     \citep[e.g.][]{ad12,ha16,qu20}.
Especially, \citet{no20} succeeded in reproducing the structural variation
of disc galaxies as a function of the galaxy mass that is revealed by the photometric decomposition 
of stellar contents into thin and thick discs and bulges
     \citep[e.g.][]{yo06,co14}.
Examinations in the present work are based on the same evolution model 
as employed in \citet{no20}, but we here concentrate on the age structures of the bulges
and compare the model with the currently available observational data.

The cold accretion theory which underlies the present study 
predicts three different regimes for the properties
of the gas distributed in the dark matter halos depending on
the virlal mass and the redshift 
     \citep[e.g.][]{de06,oc08}.
It introduces two characteristic mass scales :
$M_{\rm shock}$ above which the halo gas develops a stable shock that heats the gas
nearly to the virial temperature and
$M_{\rm stream}$ below which part of the halo gas remains cold and is confined 
into narrow filements that thread the smoothly distributed shock-heated gas.
The latter mass scale is valid only for high redshifts.

These mass scales demarcate three different regions as depicted in Fig.1.
The halo gas in Domain F is unheated and expected to accrete in free-fall to the inner region (the disc plane).
In domain G, the gas heated to the virial temperature 
attains near hydrostatic equilibruium in the halo gravitational field
and the radiative cooling induces cooling flow to the center 
with the cooling timescale.
The gas behaviour is not so clear in Domain H, where cold gas streams coexist with 
the surrounding shock-heated hot gas. 
They may behave independently and accrete with their own timesclaes 
or they may interact with each other leading to modification 
of accretion timescales. Because no detailed information is available, 
we assume that the cold and hot gases in Domain H accrete
with the free-fall time and the radiative cooling time, respectively.

In \citet{no20}, the one-to-one correspondence was assumed between the gas components in this 
diagram and the three galactic mass components of disc galaxies.
Namely, the cold gas in Domain F produces thick discs, and thin discs are formed from
the hot gas in Domain G. This part of correspondence is supported from the chemical
point of view because it 
gives a satisfactory reproduction of the stellar abundance distribution 
for the Milky Way thin and thick discs \citep{no18}.
The hot gas in Domain H is assumed to result into additional thin discs,
while the cold gas produces bulges. This whole hypothesis can reproduce the observed variation 
in the mass ratios of thin discs, thick discs, and bulges with the galaxy mass
as shown in \citet{no20}.
The cosmological simulation of \citet{br09} also suggests the correspondence between 
the gas properties and the resultant galactic components similar to the one assumed 
in \citet{no20}. 

The existence of surrounding hot gas, characteristic of Domain H, may indeed provide favourable 
condition for bulge formation from the embedded cold gas. \citet{bi16} have shown
that the external pressure resulting from AGN feedback triggers active star formation in galactic discs by 
promoting the formation of massive clumps in the destabilized discs. 
The simulation by \citet{du19} may provide another relevant result. It shows that
the ram pressure of the hot gas around massive galaxies exterted on inplunging dwarf galaxies
confines their metal-rich gas produced by supernovae and stellar winds, leading to subsequent 
star formation.
The hot halo gas in Domain H 
thus may help clumps formed in the cold gas streams survive until they create centrally concentrated stellar systems for example by radial migration. Based on these considerations,
we adopt the same correspondence hypothesis as in \citet{no20} in this study.

The locations of the borders of three domains shown in Fig.1 actually depend on the detailed physical state (e.g., metallicity, temperature) 
of the gas infalling into dark matter halos which is only poorly constrained 
by the observation
     \citep[e.g.][]{de06,oc08}.
 We use the same configuration as adopted in \citet{no20} because it leads to 
 good reproduction of the observed 
 mass fractions of the thin discs, thick discs, and bulges
 as a function of the galaxy mass observed by \citet{yo06} and \citet{co14}.
 To be specific, we assume that
$M_{\rm shock} = 1.5 \times 10^{11} M_\odot$ and
${\rm log } M_{\rm stream} = 9.2+1.067z$.
Both $M_{\rm shock}$ and
$M_{\rm stream}$ are smaller than those indicated in Fig. 7 of \citet{de06} but
closer to the \citet{ke05} shock mass and the \citet{oc08} stream mass.

The model used here treats a disc galaxy as a three-component stellar system 
comprising a thin disc, a thick disc and a bulge embedded in a dark matter halo 
that grows in mass as specified by the hierarchical mergers of dark matter halos.
Following \citet{we02}, the growth of the virial mass is given by 

\begin{equation}
	M_{\rm vir} = M_{\rm vir,0} e^{-2z/(1+z_{\rm c})}
\end{equation}
	where $M_{\rm vir,0}$ is the present halo mass and 
	$z_{\rm c}$ is the collapse redshift explained below.
The NFW density profile is assumed with the evolving concentration parameter

\begin{equation}
	 c(z) = \min \left[ K{{1+z_{\rm c}} \over {1+z}} , K \right]
\end{equation}
	with $K=3.7$ \citep{bu01}. $z_{\rm c}$ is calculated
once the present concentration $c(0)$ is specified.
We assume following \citet{ma08} that

\begin{equation}
	\log c(0) = 0.971 - 0.094 \log(M_{\rm vir,0}/[10^{12}h^{-1}{\rm M}_{\odot}])
\end{equation}

Growth of each component is driven by the accretion of gas from the halo, the timescale of which 
is determined by the cold accretion theory. 
Namely, the gas newly added to the halo in Domain F is asummed to accrete with the 
free-fall time (dynamical time) defined by $(G M_{\rm vir}/R_{\rm vir})^{1/2}$ at that moment, where $R_{\rm vir}$ is the virial radius. The accretion timescale in Domain G is the radiative cooling
time of the collisionally excited gas with the halo virial temperature and metallicty 
$Z=0.01 Z_\odot$,
calculated from \citet{su93}. The gas density is taken to be the halo density at the 
virial radius multiplied by the cosmic baryon fraction of 0.17, assuming the NFW
density profile.
 The cold gas which occupies half the newly added gas in mass in Domain 
H is assumed to accrete with the free-fall time whereas the residual hot gas accretes with 
the radiative cooling time. The gas mass added is assumed to be the increase of the halo total 
mass multiplied by the cosmic baryon fraction. These specifications determines the mass accretion rate for each gas
component completely.

We do not consider the internal structure (i.e., the density distribution) of each 
stellar component. Each component is characterized only by its mass
	and we calculate its time variation under the gas accretion from the halo.
 Actually, a significant part 
of the accreted gas is expected to escape from the galaxy due to feedback from 
star formation events such as supernova explosions especially for 
low mass galaxies. The fraction of this expelled gas
is assumed to be proportional to the inverse of the halo virial velocity at the
accretion time and the mass of the expelled gas is adjusted so that 
the stellar-to-virial mass ratio at present agrees with the observed one 
(Fig.5 of \citet{ro15} for blue galaxies).
In this study, we assume that the cold gas contained in the halo in Domain H is
turned into bulge stars immediately when it accretes onto the disc plane 
(the disc arrival time, $t_{\rm arr}$). This is likely to be 
 oversimplification and the possible effect of delay is discussed in section 4. 
 Another caveat is that it is not clear if galaxies at high redshifts have a disc.
 \citet{de20} argue that thin discs cannot develop below the critical stellar mass
 of $\sim 10^{10} M_\odot$ due to frequent mergers.  The observation by \citet{zh19} suggests that galaxies in this mass range 
 tend to be not discy but prolate at $z \sim 2$. The present model cannot discuss the shape evolution
 of galaxies by construction and the disc envisaged in the present study should not 
 be taken literally but shoud be more appropriately regarded as the inner part where most stars are distributed.

Bulge formation in the present scenario is restricted to relatively massive galaxies.
We run a series of models with the present halo virial masses  in the range
$4.33 \times 10^{11}{\rm M}_\odot \leq M_{\rm vir,0} \leq 4.98 \times 10^{12}{\rm M}_\odot$. 
The evolution of more massive galaxies 
is likely to be dominated by mergers that could turn 
 those galaxies into elliptical galaxies.
 The tracks of calculated models are shown in Fig.1. The least massive model 
 (and models less massive than this) does not 
 exter Domain H so that no bulge component is formed.

\section{Results}

Fig.2 illustrates the star formation history for each model.
It is seen that more massive models form bulges earlier than less massive ones
and the bulge formation in those models spans longer periods in time.
These trends are illustrated in a different form in Fig.1, where two black circles on 
each evolutionary track indicate the redshifts at which the cold accretion 
originating in Domain H reaches to the disc plane first and last.
This mass dependence of bulge formation history is quantified and compared with observations later.

\begin{figure}
	\includegraphics[width=1.0\linewidth]{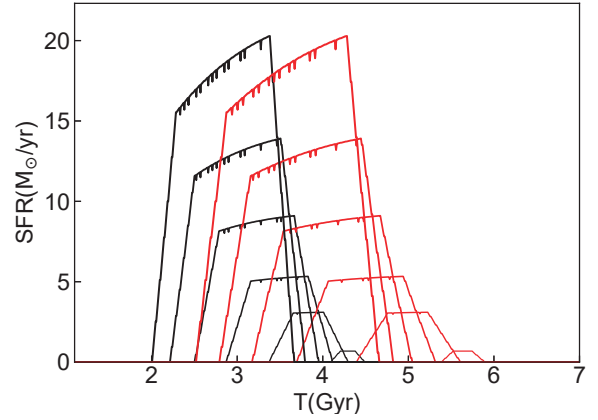}
    \caption{
	    Star formation rate for the bulge component as a function of time,
	    with thicker black lines indicating models with larger virial masses  
	    at present. 
	    Plotted values are running means with the width of
	    0.28 Gyr. Tiny spikes are caused by the numerical method used in the evolution
	    model and do not affect our conclusions. 
	    Red lines indicate the star formation history for which delay of
	    twenty dynamical times is taken into account.
         }
    \label{Fig.2}
\end{figure}

\begin{figure*}
	\includegraphics[scale=0.3]{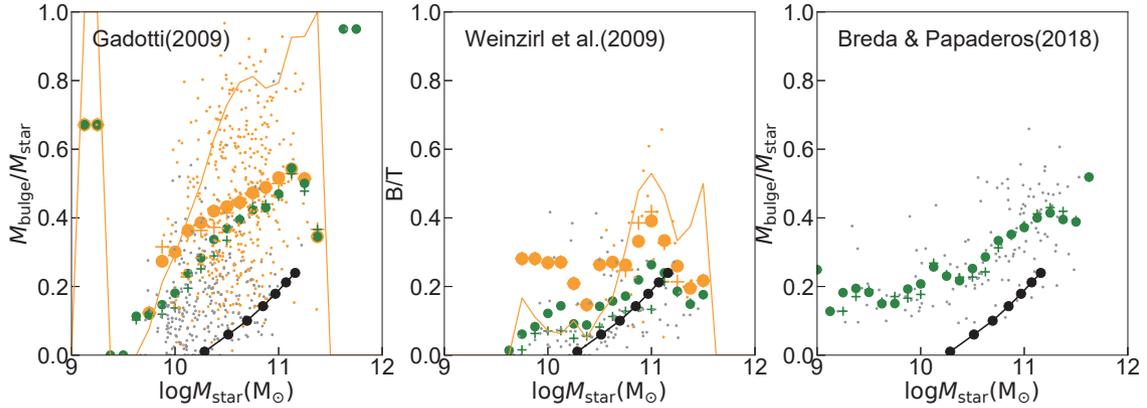}
    \caption{
	    Model bulge fractions compared with three sets of observations.
	    Black dots connected by solid lines are model results, whereas
	    observational data are represented by small dots.
	    Green circles and pluses indicate, respectively, the running mean and median 
	    in the mass bin
	    having the width of 0.25 dex and moved by every 0.125 dex in the galaxy total stellar mass.
	    In the left and central panels, orange symbols denote means and medians only for classical
	    bulges (orange small dots) defined to have the Sersic index larger than 2 
	    in i-band and H-band, respectively.
	    The orange lines indicate the number fraction of classic bulges in each mass bin.
	    \citet{br18} do not derive the Sersic index and no classification of bulges is
	    possible.
         }
    \label{Fig.3}
\end{figure*}

\begin{figure}
	\includegraphics[scale=0.4]{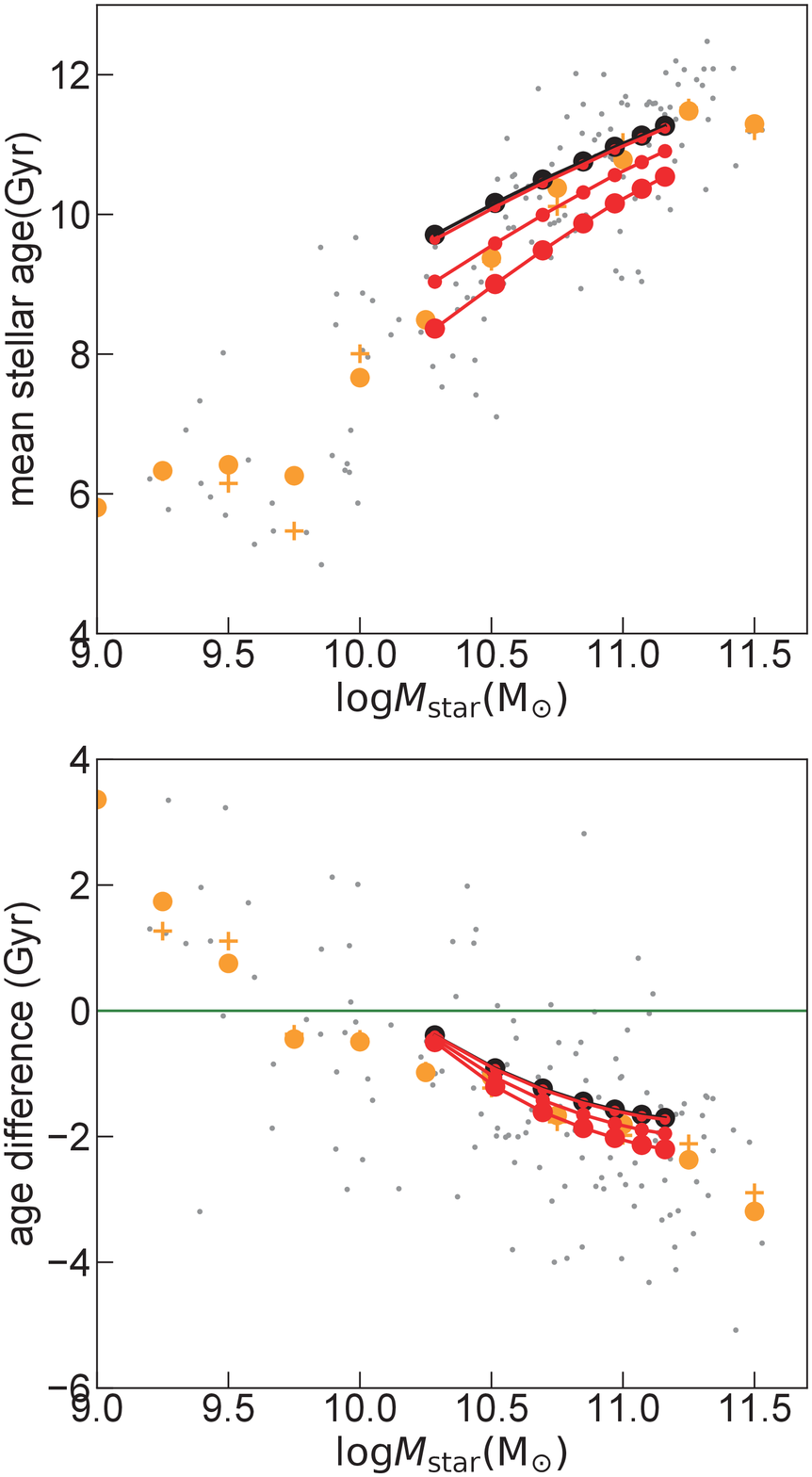}
    \caption{
	    The mass-weighted mean stellar age of the bulge (upper panel) 
	    and the age difference over the bulge radius (bottom panel) 
	    are compared with the observational data of \citet{br18} and
	    \citet{br20}, respectively.
	     Each observed value is plotted with gray. Orange circles and pluses
	    indicate the mean and median in each mass bin with the width of 0.25 dex.
	    No star formation delay is taken into account for black circles.
	Red circles indicate, in increasing size, the results with the delay time,
	$t_{\rm dyn}, 10t_{\rm dyn}$,
	and $20t_{\rm dyn}$.
	}
    \label{Fig.4}
\end{figure}

Fig.3 shows the mass fraction of the bulge as a function of the total stellar mass of the 
galaxy at the present epoch. 
We compare the model with three sets of the observation. 
The sample of \citet{ga09} comprises nearly face-on galaxies extracted fron the Sloan 
Digital Sky Survey.
\citet{we09} decomposed H-band images of 
 S0/a-Sm galaxies in the Ohio State University Bright Spiral Galaxy Survey \citep{es02}.
 Finally, \citet{br18} analyzed 135 late-type galaxies from the CALIFA survey \citep{sa12,sa16}.
The observed bulge masses
likely suffer from large uncertainties as guessed from this figure. Nevertheless, 
the three different analyses consistently indicate the increase in bulge mass fraction
with the total stellar mass. The model reproduces this qualitative trend.
Model values agree well with the result of \citet{we09} but about half the values 
reported by \citet{ga09} and \citet{br18}.
We discuss possible reasons for this discrepancy later.

Fig.4 summarizes the age structures of the model bulges. 
The mean stellar age plotted in the upper panel increases with the total stellar mass
in qualitative agreement with the observation by \citet{br18}. 
However, the model dependence (black circles) is shallower than the observed one and the discrepancy 
increases toward lower galaxy masses. 
We discuss later the effect of including 
possible delay in the bulge star formation. 
The present model predicts the lower mass limit 
for bulge formation
originating in Domain H around 
$M_{\rm star} \sim 10^{10}{\rm M}_\odot$. 
We discuss later possible different origins for bulges in lower mass galaxies 
plotted in Fig.4.

The age difference plotted in the bottom panel is simply the time at which the
bulge star formation starts minus the time at which it ends. Because the gas that accretes
to the disc plane later ends up at a larger distance from the galactic center,
the age difference thus defined essentially corresponds to the 'age gradient within 
the bulge radius' shown in 
Fig.2 of \citet{br20}. It should be noted that the 'gradient' given in \citet{br20}
is not the age difference per unit length but the difference 
between the outer and inner edges of the bulge.
Reflecting the inside-out nature of gas accretion, all the models 
produce negative gradients. 
Furthermore  the absolute value of the gradient increases with the galaxy 
mass. Over the mass range for which the model produces bulges, the model values 
are in good agreement with the observed ones.

\section{discussion and conclusions}

We have shown that the present model which is based on the cold accretion scenario 
for galactic gas accretion reproduces the 
observed bulge properties despite its idealized nature although some discrepany remains.
In alliance with its success in explaining the chemical bimodality in the Milky Way 
disc \citep{no18} and the morphological variation with the galaxy mass 
observed for extra-galaxies \citep{no20},
this result may be regarded to reinforce the cold accretion scenario from the viewpoint 
of internal structures of individual disc galaxies.
Nevertheless, there are missing ingredients in the simplified approach taken here.
We touch upon these unresolved issues in the following.

In addition to the bulge mass fraction, the bulge size is also known to increase with 
the galaxy total stellar mass
 \citep[e.g.][]{ga09}.
Although the present model cannot determine the bulge size
because of its one-zone nature, it may be instructive to make rough estimate 
for the expected size from the virial radius $R_{\rm vir}$ and the 
spin parameter $\lambda$ of the dark matter halo.
The size calculated as $r_{\rm bulge} = \lambda R_{\rm vir}$ at the bulge formation
epoch is $2 \sim 3 $ kpc assuming $\lambda=0.03$, which is similar to the obseved sizes 
for the most massive galaxies but depends little on the galaxy mass
for the calculated mass range.
This is because the lower-mass galaxies experience bulge formation later than 
the higher-mass galaxies so that the virial radius at the bulge formation epoch 
as defined in this study
is nearly constant with the galaxy mass.

	We assumed that the cold gas in Domain H is turned into stars upon its arrival
at the disc plane. This assumption may be oversimplified. It is conceivable that
the cold gas streams contain gas clumps and after disc arrival individual clumps are transported inward due to
violent disc instability (VDI) before star formation occurs in them (or while making stars en route
to the galactic center). Clump formation within the cold gas filaments due to 
gravitational instability is suggested by \citet{ma18} in relation to globular cluster
formation. Many cosmological simulations also reveal gas clumps
in those filaments 
 \citep[e.g.][]{ke05,de06,oc08,va12,ne13}, part of which are brought in to forming
 galactic disks \citep{ds09}.
 Radial migration timescale due to VDI is estimated to be of the order of ten times 
 the dynamical time \citep{de14}. Red circles in Fig.1 and Fig.4 illustrate how this delay affects 
 the star formation epochs and age structures of the bulges. We see that the inclusion 
 of star formation delay improves the agreement with the observation (especially bulge ages) with the delay time 
 of ten times the dynamical time giving the best fit.
 Influence of delay is larger for smaller galaxies because of longer migration times,
 resulting to significantly younger bulge ages than the fiducial case (black circles 
 in Fig.4).

	Bulge formation in the present study may be related to 
the compaction and blue nuggets (BNs) reported in the cosmological simulation by \citet{zo15}.
Fig.1 plots the simulated compaction events on the $z-M_{\rm vir}$ plane.
They are located in the bulge formation region 
in the present model bordered by black and red circles.
The peak masses for the BNs observed by \citet{hu18} in different redshift ranges also fall
on the domain expected for bulge formation once a certain delay from the disc arrival 
is taken into account. 
The star forming galaxies at $z\sim2$ observed by \citet{ta16}  
exhibit different star formation profiles depending on the stellar mass with galaxies
of intermediate masses 
($10^{10.1} {\rm M}_\odot < M_{\rm star} < 10^{10.6} {\rm M}_\odot$) showing more 
centrally-concentrated profiles than either less massive or more massive galaxies.
This result also seems to be in line with the present study which proposes bulge 
formation in the restricted mass range, 
$M_{\rm shock} < M_{\rm vir} < M_{\rm stream}$.

The present model predicts bulge formation only above a certain threshold for 
the present galaxy mass
around $M_{\rm star} \sim 10^{10}{\rm M}_\odot$. 
It is possible that bulge formation involves several mechanisms and those bulges 
in less massive galaxies are formed by different processes. One possibility is 
the secondary bulge formation from disc material in later cosmological epochs
as mentioned in introduction.
Indeed, the upper panel of Fig.4 shows a steep decrease in bulge ages below 
$M_{\rm star} \sim 10^{10}{\rm M}_\odot$ in the observation by \citet{br20}.
The age gradient (the bottom panel) also turns to positive below this critical mass, suggesting
a different mechanism operating other than the inside-out gas accretion form the halo.
There seems to be a tendency that classical bulges inhabit massive galaxies whereas 
pseudo bulges are observed in less massive galaxies \citep{ga09,we09,fi11}.
This habitat segregation may make the cold-accretion driven bulge formation 
proposed in this study a likely candidate specific to classical bulge formation.
Indeed, Fig.3 shows that the threshold mass for bulge formation in the model nearly
coincides with the mass above which the classical bulges start to emerge in actual disc galaxies.
If this inference is correct, part of the discrepancy between the model and observations
appearing in Fig.3 may be also solved. The observed excess of the bulge mass in 
\citet{ga09} and \citet{br18} could be contributed by secular processes.
On the other hand, the bulge fraction in \citet{we09}, which is actually the luminosity
fraction in H-band, could be underestimated if the stellar population in the bulge 
is systematically older (and threfore redder) than the disc in those galaxies, which is 
quite likely.
	In either case, we need not consider that the classical and pseud bulge formation
	processes occur excluisively with each other. Regading bulge formation, 
	the galaxy mass sequence may be a continuous sequence along which the relative 
	importance of two (or more) bulge formation processes changes gradually.

We applied for the first time the cold-accretion driven galaxy evolution model
to the currently available observational data for bulge properties
in galaxies with various masses. The model, despite its highly idealized nature,
can reproduce the observed behaviours at least qualitatively, although 
observational data are still meager and future observations are required 
to construct a more concrete picture for bulge formation.
Especially, galaxies at $z \sim 2-3$ will provide wealth of information on 
the bulge formation because galactic bulges are thought to grow vigorously 
in this epoch (see Fig.1). It is interesting that \citet{ta16} found a sign for 
increasing bulge dominance for more massive galaxies in this redshift range
 in agreement with the theoretical result by \citet{no20}.
The scrutinization of internal properties of the nearby bulges such as performed by
\citet{br18} and \citet{br20} will put constraints at the present cosmological epoch,
playing a complementary role with high-redshift surveys.

On the theoretical side, recent cosmological simulations start to produce disc galaxies
with realistic bulge-to-disc mass ratios unlike early simulations that produced too massive bulge components
     \citep[e.g.][]{ma14,ga19}. \citet{ga19} report that their bulges in the Auriga simulation
     comprise mostly in-Situ stars and merger contribution is negligible.
The work of \citet{br09} is pioneering in that it related different structural components
of disc galaxies formed in the cosmological simulations to different modes of
gas accretion, namely accretion of clumpy, shocked, and unshocked gas. Although high-resolution cosmological simulations are
very expensive, such close inspection of even a small number of created galaxies 
will provide valuable insight into the build-up of disc galaxies free from 
idealization made in the present work.

\section*{Acknowledgements}

We thank Iris Breda and Polychronis Papaderos for providing the observational data for galactic bulges and stimulating discussion on the bulge formation mechanisms.
We also thank the anonymous referees for invaluable comments which helped improve the 
manuscript.

\vspace{16pt}

\noindent{Data availability}

\vspace{6pt}
\noindent{The data underlying this article will be shared on reasonable request to the corresponding author.}






\bsp	
\label{lastpage}
\end{document}